\documentclass[10pt]{article}
 
 \usepackage[scaled]{helvet}
 \usepackage{authblk}
\usepackage{tikz}
\usepackage{url}

\usepackage[T1]{fontenc}
 
\usepackage{fullpage}
\usepackage{amssymb}
\usepackage{amsmath}
\usepackage{setspace}
\usepackage{graphicx}
\usepackage{subfigure}
\usepackage{algorithm}
\usepackage{algorithmic}
\usepackage{color}
\usepackage{qtree}

\newcommand{\die}[2]{\frac{\partial #1}{\partial #2}}
\newcommand{\dee}[2]{\frac{d #1}{d #2}}

\title{PopED lite: \\ an optimal design software \\ for preclinical pharmacokinetic and pharmacodynamic studies}
\author[1]{Yasunori Aoki \thanks{Corresponding author: yaoki@uwaterloo.ca}}
\author[2]{Monika Sundqvist}
\author[1]{Andrew C. Hooker}
\author[2]{Peter Gennemark}
\affil[1]{Pharmacometrics Research Group, Dept. Pharmaceutical Biosciences, Uppsala University, Box 591, SE-751 24, Uppsala, Sweden}
\affil[2]{CVMD iMed DMPK AstraZeneca R\&D, SE-431 83 M\"{o}lndal, Sweden}

\begin{document}  
\maketitle
\begin{doublespace}
\begin{abstract}
Optimal experimental design approaches are seldom used in pre-clinical drug discovery. Main reasons for this lack of use are that available software tools require relatively high insight in optimal design theory, and that the design-execution cycle of in vivo experiments is short, making time-consuming optimizations infeasible. 
We present the publicly available software PopED lite in order to increase the use of optimal design in pre-clinical drug discovery. PopED lite is designed to be simple, fast and intuitive. Simple, to give many users access to basic optimal design calculations. Fast, to fit the short design-execution cycle and allow interactive experimental design (test one design, discuss proposed design, test another design, etc). Intuitive, so that the input to and output from the software can easily be understood by users without knowledge of the theory of optimal design. In this way, PopED lite is highly useful in practice and complements existing tools. 
Key functionality of PopED lite is demonstrated by three case studies from real drug discovery projects.

\end{abstract}

\section{Introduction}

Poorly designed experiments often lead to uninformative experimental results. For example, poorly chosen dosage and observation/sampling times in an animal experiment can lead to inaccurate characterisation of drug candidates. In order to minimize the risk of running uninformative experiments in drug discovery, we introduce the software PopED lite that systematically incorporates a priori knowledge about the compound and algorithmically optimizes the experimental design.

Decisions in preclinical drug discovery are largely based on inference of data from cell-based assays and animals exposed to drug candidates. Accurate efficacy and safety estimates of each test compound are pivotal for proper compound ranking. Mathematical models are commonly used to formally integrate available information in order to gain system-level understanding of pharmacological effects and to make informed decisions~\cite{Lalonde2007, Visser2013}. These models are usually composed of two parts; the pharmacokinetics (PK) part representing what the body does to the drug, and the pharmacodynamics (PD) part representing what the drug does to the body~\cite{GabrielssonJ2010}.% (Appendix \ref{se:appendixModels}).

During the compound selection phase of a drug discovery project, several compounds are tested. Estimated model parameters (e.g., clearance and potency) are used to quantitatively characterise the compound. Naturally, the uncertainty of parameter estimates must be taken into account. The design of the experiment (e.g., when to take the blood samples, how much compound to administer) can have large influence on the parameter estimation uncertainty. Therefore, experimental design of preclinical studies is a crucial component of overall success in drug discovery.

Currently, in a typical preclinical study, experiments are manually designed and sampling times are often fixed for all test compounds. Although there have been significant developments in optimal design for clinical studies~\cite{Fedorov2007,Lledo-Garcia2012, Mentre1997, Nyberg2009, Nyberg2012}, the use of optimal design in preclinical studies is still limited~\cite{Sjogren2011, Sjogren2012}. Thus, we hypothesise that there is significant room for improvement in the experimental design of  preclinical studies. For example, the power of prior information from similar compounds is not fully taken into account.

Initially, we attempted to introduce already existing optimal design software~\cite{Bazzoli2010, Nyberg2012} into a realistic experimental design workflow of pre-clinical PK/PD studies. During this attempt, we learned how preclinical experiments were designed in drug discovery and identified several causes for optimal design not being used in pre-clinical studies. Specifically, we identified three key challenges in utilizing the idea of optimal design in preclinical experiments. 

Firstly, the currently available experimental design tools are developed by academic groups with the aim of delivering state of the art optimal design theories to a wide variety of applications.  Therefore these software tools are developed for high functionality and high flexibility.  However, flexibility incurs more user inputs which can demotivate unexperienced users and increase the risk of user errors.  As the development of experimental designs for similar preclinical studies is often done routinely, the high flexibility is not necessary and hinders users who are not familiar with optimal design software tools.

Secondly, most of the currently available software tools are designed to be used for clinical studies.
For a clinical study the true optimality of the design is desirable, since study cost  is high and since the design-experiment cycles are in the order of months.   
As a consequence, currently available software tools sacrifice speed of computation for optimality of the design.  However, for preclinical studies, the design-experiment cycles are in the order of weeks, and often the experimental designs are adjusted during discussions with experimentalists.  Thus, an optimal design software that requires considerable computational time is not suitable for preclinical studies.

Thirdly, most of the optimal design software tools assume basic user knowledge in optimal design, and use optimal design terminology (e.g., D-optimal criterion, determinant of FIM,...) in the user interface to concisely communicate the idea behind the software.  
%On the other hand, as optimal design in preclinical studies are still a new idea, it is not possible to expect the preclinical experimentalists to be familiour with optimal design terminologies.  
 However, the majority of preclinical experimentalists are not used to optimal design terminology and interpretation of the required user-input and software-output of these programs becomes an issue.

To address these three challenges we have implemented the software PopED lite that is designed to be a simple, fast and intuitive playground for experimental design in the preclinical area of drug development.  PopED lite is available on the Mac App Store \url{https://itunes.apple.com/se/app/poped-lite/id836277613?l=en&mt=12} for Mac and at \url{http://www.bluetree.me/PopED_lite} for Windows.

\section{Design Considerations}

%Optimal experimental design is useful to maximize information in data from an experiment and minimize the experimental resources (e.g. animals and time). PopED lite aims to bring the idea of optimal design into the regular workflow of preclinical animal experiments in the pharmaceutical industry.

Through our initial attempt of introducing pre-existing optimal design software tools to industrial drug-discovery teams, we observed that these were too complex for the relatively simple experimental design computations needed for preclinical PK/PD studies, too slow to fit into the compound-selection workflow, and too difficult to understand for users without prior knowledge in optimal design.  PopED lite was designed to be simpler, faster and more intuitive compared to existing programs in order to bring optimal design into the regular workflow of preclinical animal experiments in the pharmaceutical industry.
%To design PopED lite, we first assessed the necessary flexibility of the software for use in preclinical experiment so that the software is as simple as possible. To avoid unnecessary precise optimization of the design, we also carefully investigated the level of precision needed for practical use. Then, lastly, we considered the background technical knowledge of a typical industrial drug-discovery team so that the interface of the software is intuitive to users.
Together with an industrial drug-discovery team we conducted a retrospective study (Case Study 1) and with another team we designed experiments using a preliminary version of PopED lite (Case study 2) as well as with the current version (Case Study 3). Through these case studies in addition to the discussions with many other preclinical experimental teams, we have attempted to balance the appropriate flexibility, accuracy and level of technicality.

\textbf{Flexibility}.  
PopED lite focuses on optimization of the dosage and PK/PD sampling (observation) times to improve the accuracy of the parameter estimates of fixed effect PK/PD models.  Pre-clinical experiments are characterised by relatively small group sizes, and animals with small genetic differences. The main interest is to predict the human situation, and in this translation focus is on average levels and not on variability, as translation of the latter is associated with very high uncertainty. Therefore, the use of nonlinear mixed effect modelling is not as central as in the clinical setting, and key questions can often be addressed by standard deterministic (fixed effect) PK/PD models. 

PopED lite implements standard compartmental PK models (1-, 2-, and 3-compartment models with linear and nonlinear elimination, and linear absorption), as well as commonly used PD models. For PD models, the user can alternatively input a regular mathematical expression in form of a function or an ordinarily differential equation. This choice of model flexibility is chosen to cover a wide range of possible PK/PD models and at the same time ensure software usability.

%In addition, we have also observed that standard 1-, 2-, and 3-compartment PK models with linear and nonlinear elimination, and linear absorption are used in the majority of preclinical PK/PD studies.  On the other hand, we have observed many variations of the PD models, so commonly used PD models are implemented for convenience; however, the user can define the PD model by regular mathematical expressions in form of a function or an ordinarily differential equation.

%Sampling time and dosage are used as experimental design variables; these are considered most important in preclinical experiments.

\textbf{Accuracy}. 
We have decided to discretize both sampling time and dosage search space, so that the optimal design can be chosen from a combination of practical sampling schedules and allowed possible dosages.

We have observed that a precisely optimized experimental design is seldom practically useful. For example, in practice one can only sample every 5th minute and a discrete search space for sampling or dose times is sufficient. By discretizing the sampling time and dosage the design space will shrink to finite, and we can reduce the computational cost for the design optimization while avoiding unrealistically precise experimental designs.

We have also observed that a quickly obtained good-enough design, and not necessarily the optimal, is often more valuable in preclinical studies due to their time constraints.

\textbf{Level of Technicality}. 
Preclinical studies are designed by cross-functional teams (e.g., biologists, chemists, and PK/PD-modellers), that generally lack training in numerical computation and information theory. Therefore, optimal-design specific jargon should be avoided in the software interface, and PopED is designed to only display terminology that is familiar to these teams. Then based on the information specified by the user, PopED lite chooses an appropriate design optimization strategy.
In this way, PopED lite introduces the idea of optimal design to preclinical experimentalists. 

%Furthermore, the initial setup of the software should be as easy as the installation of any other desktop applications (Apps).

In summary, the design philosophy of PopED lite is to create the simplest, fastest, most intuitive optimal design software that maintains the appropriate flexibility, accuracy and level of technicality for use in preclinical drug discovery.

\section{Material and Methods}
In PopED lite, we obtain an optimized experimental design by solving a constrained optimization problem, where each experimental design is evaluated using some function of the Fisher Information Matrix (FIM).  The FIM is calculated based on the PK/PD model structure, possible sets of model parameters, and the experimental design (see \cite{AtkinsonAC1992} for an introduction to the field).

\subsection{Overall Workflow and User Interface}
\begin{figure}
\centering
\includegraphics[bb=0 0 500 829, scale=0.69]{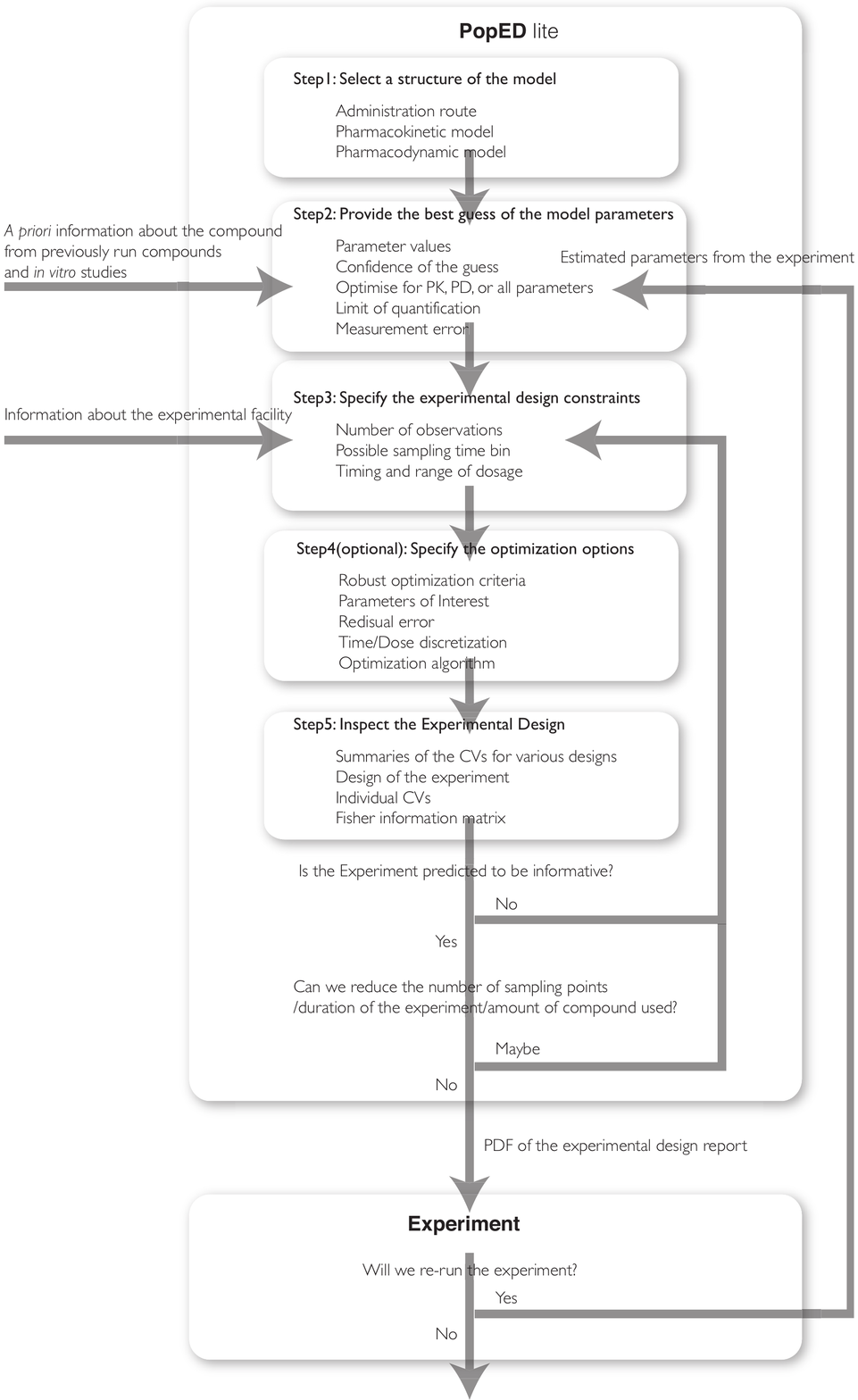}
\caption{Overall workflow of PopED lite.}\label{fig::workflow}
\end{figure}

User input to PopED lite is entered step by step; completion of one step is required before the next step appears. This increases usability and reduces user errors.
The optimal design is calculated in five steps (Figure~\ref{fig::workflow}). Users are only required to input numbers, and a single mathematical expression in case a user-specified PD model is desired.
%This ensures the causality of the workflow, makes the software easy to use, and minimizes the risk of user errors.
%The optimal experimental design is typically calculated in the following way: the goodness of each design from the design space is quantified by some measure (the optimization criteria) of the Fisher Information Matrix (FIM). The FIM contains information of how sensitive the model is to changes in parameters, as well as information about the error model. The best design is output.

\subsection{Computation related to the Fisher Information Matrix}
Accurate construction and computation of the FIM is vital to optimal design software, as the FIM is used to quantify the quality of the experimental design. In order to assure appropriate FIM construction, the residual error model is automatically built using the estimated limit of quantifications and measurement errors provided by the user in Step 2 (Figure~\ref{fig::workflow}) of PopED lite. In addition, since the FIM is a symmetric matrix PopED lite utilizes Cholesky decomposition to calculate the determinant and inverse of the FIM. At each Cholesky decomposition, PopED lite monitors the rounding error and also checks the invertibility of the FIM. If the FIM of the optimized design is singular, the program outputs a warning message saying that the parameters cannot be uniquely identified from the experiment, instead of stating that the FIM is singular. Naturally, this facilitates user interpretation.
In addition, if the computational error is detected when inverting the FIM, PopED lite outputs a warning as well as reports the corresponding Coefficients of Variation (CVs) with a question mark (?), allowing the user to interpret these CVs with caution. Appendix~\ref{se:appendixFIM} describes in detail how the FIM is constructed in PopED lite.

\subsection{PK/PD model simulation}
The majority of the computational cost for optimal design software originates from the computation of the FIM. This computational cost in calculating the FIM is reduced by hard-coding several PK and PD models in PopED lite.
%%FIM calculations are typically fast in PopED lite since common PK/PD models and their derivatives are hardcoded (compiled C++ implementation). 
For user defined PD functions, PopED lite implements an expression parser based on the shunting-yard algorithm \cite{shuntingyard}. PopED lite parses the expression only once to save computation time. All subsequent evaluations of the expression in the design optimization procedure reuse the parsed expression.  For PK/PD models that require solving a system of ordinary differential equations, we use the Rosenbrock method \cite{Rosenbrock1963} implemented in the boost package in the Odint toolbox \cite{Ahnert2011}.  The Jacobian of the right hand side of a user defined PD function is approximated using a forward difference scheme.

\subsection{Optimization Criteria}
PopED lite chooses the optimization criteria from D, Ds, ED, and EDs optimal designs~\cite{AtkinsonAC1992, Fedorov1972} to fit the optimization need of the user.
For a chosen PK/PD model structure, the user can specify the confidence level of the best guess of the PK/PD parameters (Step 2 of PopED lite). If there is no uncertainty, PopED lite optimizes the experimental design using the D optimization criterion. Otherwise, it randomly generates sets of parameters following the uniform distribution in the range specified by the user and conducts ED optimal design (by taking the expectation of the natural logarithm of the determinant of the FIM for randomly generated sets of parameters).
Furthermore, the user can specify in Step 2 that PopED lite should optimize the design for parameter estimation accuracy of a subset of parameters; Ds or EDs optimization criteria will then be used.  Note that these optimization criteria do not appear in the standard user interface and PopED lite interprets the optimization needs of the user and chooses a criterion automatically.

\subsection{Optimization Algorithms}
We have implemented discrete optimization algorithms, in order to achieve the desired speed and also to obtain an experimental design that makes sense in practice.

By discretizing the sampling time, it is possible to reduce the search space (the design space) significantly. A key advantage of discretizing the sampling time is that model simulations are avoided during sampling time optimization since all the elements that can appear in the Jacobian matrices can be pre-calculated. This dramatically reduces the cost of computation especially when the model is defined by a system of ordinary differential equations (ODEs).

By discretizing the dosage, the user can specify the possible doses considering the experimental constraints and compound formulation constraints. As a result, in particular for single dose experiments, PopED lite can search exhaustively for the dosage and hence avoid the risk of finding a locally optimal design.

The sampling time optimization is a combinatorial optimization problem (e.g. choose 6 sampling points from 420 possible sampling points), while the dose optimization is a simple discrete optimization problem. PopED lite implements stochastic optimization algorithms for these two problems to reduce the risk of finding a local optimal design. The two algorithms are detailed in Appendix~\ref{se:optAlgorithm}.

\subsection{Implementation Framework}
PopED lite was implemented using the Qt framework \cite{Digia2014} and overall implementation was optimized using Instruments \cite{Apple2014}.  The Qt framework was chosen, for its speed of computation, ease of deployment, and flexibility of GUI design.  The Qt framework is based on C++, hence PopED lite takes advantage of the speed of compiled software.  Most of the optimal design programs targeted for pharmaceutical drug development are written in interpreter based languages such as R and Matlab which is a major cause of slow computation.  Typically one order of magnitude faster computation can be expected for compiler based language such as C++, C, or FORTRAN compared to interpreter based languages.  
%In addition, to take the full advantage of the speed of complied software, the overall implementation is optimised using a profiling software Instruments \cite{Apple2014}.
The Qt is also a cross-platform framework so that PopED lite can be deployed to both Unix-based platforms and Windows-based platforms.  We have also made PopED lite a standalone software to facilitate installation.  As a result, PopED lite does not require any extra installation step and can be started by simply clicking on an icon just like any other Apps available on a computer.
Lastly, the Qt framework supports flexible GUI design with a unified look and feel.
%Lastly, the Qt framework has sufficient flexibility on the GUI design to let us design easy to use user interface (UI) with minimal UI related implementation.  As an additional benefit, the GUI designed using Qt gives an unified look and feel with other native applications thus users without high technical skills.

\section{Case Studies}
We now present the case studies that PopED lite was motivated by, developed for, and used in.
All case studies are based on problems encountered in authentic drug discovery projects.

\subsection{Case study 1: Designing a mouse receptor-occupancy study to estimate test-compound potency}
\begin{figure}
\subfigure[PK (left) and PD (right) simulation of the manually designed experiment for Case study 1.  Red dots represent the sampling time.  Red lines are the simulations with random parameters generated from the parameter guesstimates and their uncertainties.  The shaded area is the 68\% prediction interval of the observation based on the measurement error, LOQ, and parameter guesstimate.]{\includegraphics[scale=0.33]{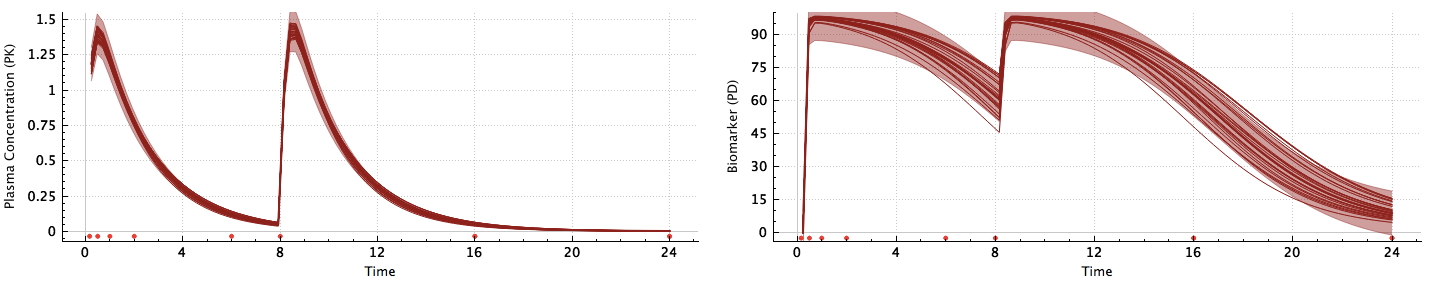}\label{fig::study1_upper}}
\subfigure[Graphical user interface of PopED lite.  Design0 in the table (the first line) is the manual design and Design1 is the design optimized by PopED lite.]{\includegraphics[scale=0.33]{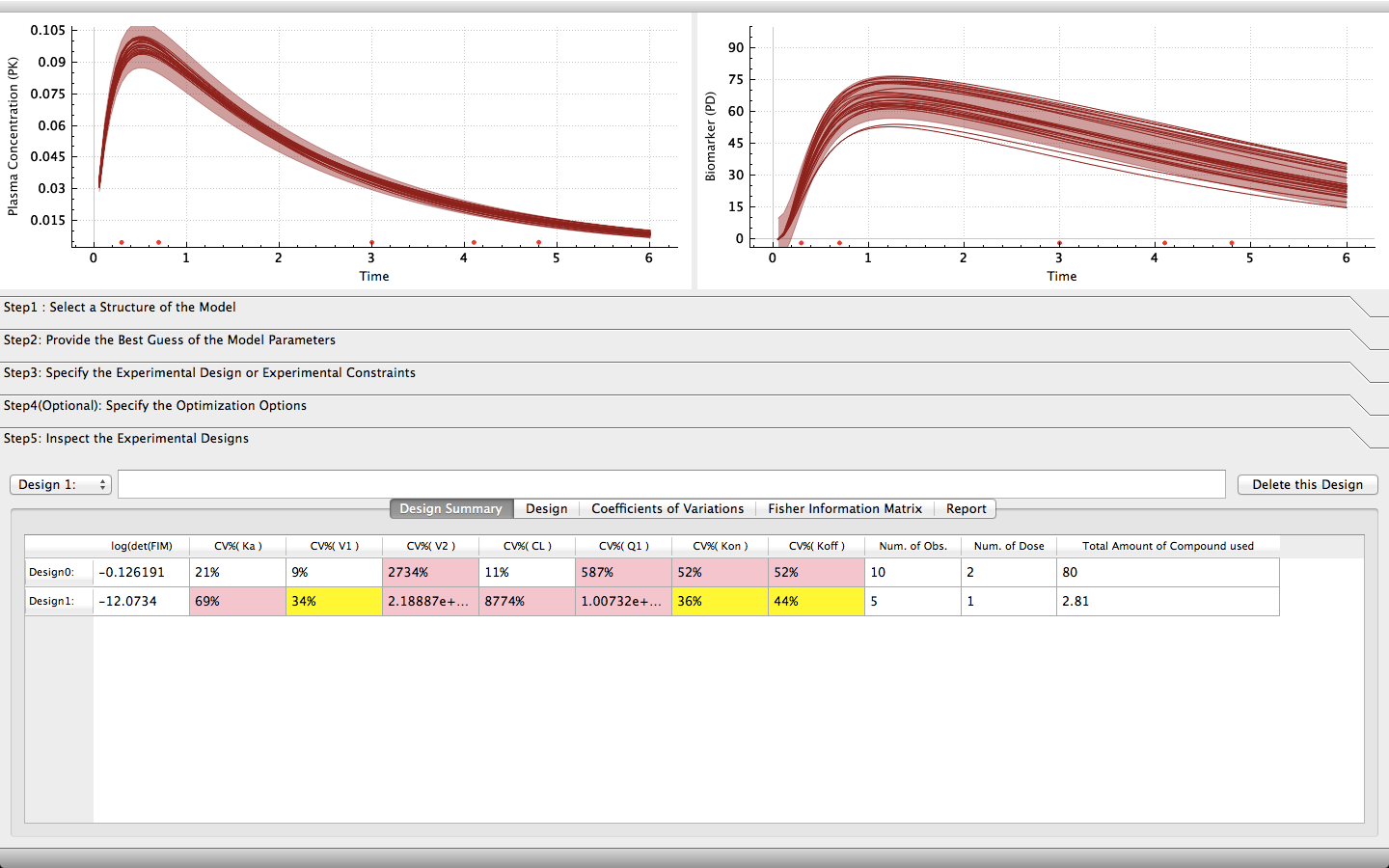}\label{fig::study1_lower}}
\caption{Case study 1. PopED lite screenshot of the summary of the experimental designs.  Design1 gives lower CV\% for $k_\textrm{off}$ and $k_\textrm{on}$ and requires less number of observations and amount of compound compared to Design0.}
\end{figure}

This case study was run before the start of the development of PopED lite, and motivated the development of optimal design software specifically designed for preclinical studies.
In a drug candidate optimization program, several chemically similar compounds were evaluated in the mouse. Specifically, time series data for drug exposure and drug targeted receptor occupancy were generated from orally dosed mice. For the latter time-series, one mouse was sacrificed for each sampling point making experimental design highly important both from ethical and cost perspectives. For the compound series, PK could be reliably modeled by a linear two-compartment model (1st order absorption, linear elimination %; see Appendix~\ref{se:appendixModels}
), and the distribution between drug-receptor complex ($RC$) and free receptor ($R$) was modeled by a receptor kinetic model with elementary reactions as
\begin{eqnarray*}
\dee{}{t}RC(t)=k_\textrm{on} C(t) (R_\textrm{tot}-RC(t))-k_\textrm{off}RC(t)
\end{eqnarray*}
where $C$ denotes drug plasma concentration, $R_\textrm{tot}$ denotes the total receptor concentration, $RC$ denotes the drug-receptor complex and $k_\textrm{on}$ and $k_\textrm{off}$ are kinetic parameters. The dissociation constant is derived as $K_M = k_\textrm{off} / k_\textrm{on}$, and reflects the potency of the test compound.
The in vivo experiment was routinely conducted using a manually defined standard design. For one test compound, the CV of $k_\textrm{off}$ was as high as 80\% resulting in high uncertainty in the potency estimate. The reason for this high CV was that all observations after 16 hr were below the limit of quantification and hence not informative. To improve the estimate, a second experiment was planned. During the experimental planning, various simulations were run with manually designed experiments and the experimental design with two doses of 40 $\mu$mol/kg (at time 0 h and 8 h) and sampling at 8, 16, 16, 24, and 24 hrs was chosen for the second experiment (Figure~\ref{fig::study1_upper}). By combining the first and the second experiments, the predicted CV for $k_\textrm{off}$ was 52\%. Actual data generated using this design reduced the CV of $k_\textrm{off}$ even more, down to 22\%, and the potency could be reliably derived.

%This example illustrates the need for automation of the design optimization process and made us aware of the lack of accessible software for preclinical studies. 

In a retrospective analysis we explain how this experiment could have been optimized even further using PopED lite.
We first tried to optimize for the sampling schedule only, but did not find significant improvement on the predicted CV for $k_\textrm{on}$ and $k_\textrm{off}$. In a subsequent step we optimized for the dosing amount using an optimization criterion taking the accuracy of all the model parameters into account (ED optimal design). The proposed design also failed to improve the CV for $k_\textrm{off}$ and $k_\textrm{on}$. However, by optimizing only for the PD parameters (EDs optimal design), the predicted CVs for $k_\textrm{off}$ and $k_\textrm{on}$ were reduced to 36 and 44\%. As the predicted CVs of $k_\textrm{off}$ and $k_\textrm{on}$ for the actual experiment ran were 52\% and 52\% (see Figure~\ref{fig::study1_lower} design 0), we can expect further reduction in the parameter estimation uncertainty with this design. In addition, this optimal design requires a short sampling period (0.3, 0.7, 3, 4.1, and 4.8 hours), and only one dosing of 2.8 $\mu$mol/kg at time 0. 

\textcolor{black}{By using EDs optimal design with PD parameters as the parameters of interest, it is not unexpected that the CVs of the PK parameters are relatively high in the proposed design. However, even if individual PK parameters cannot be identified accurately, the shape of the PK profile can be roughly estimated from the observation, and we can estimate PD parameters accurately using a relatively sparse sampling schedule.}

%Thus, by using PopED lite, we could potentially not only have reduced the parameter estimation uncertainty, but also reduced the required time of experiment, as well as the amount of compound used in the experiment.

In summary, the retrospective analysis indicates that by using 
%PopED lite  
an optimal design software we
could have reduced the required amount of compound, which is often limited in preclinical studies, and study duration while obtaining similarly accurate parameter estimates. This example also illustrates the importance of both dose and sampling time optimization as well as optimization focused on the parameters of interest (i.e. EDs optimal design).

\subsection{Case study 2: Designing a dog study to estimate test compound potency}
PopED lite was developed along with a drug discovery project and a preliminary version of PopED lite was used to design several dog experiments.
The drug discovery project aimed at decreasing the level of a biomarker by more than 80\% over 24 h. The compound concentration and biomarker response could be measured both in vitro and in vivo. However, several compounds displayed a time delay between concentration and effect in vivo that was not fully captured in the in vitro assay. To better understand the exposure-response relationship, the compound was administered to dogs in vivo, and exposure and biomarker was followed over time. The PK was then modeled using a 2-compartment model %(Appendix~\ref{se:appendixModels})
. The biomarker was modeled using a link model \eqref{eq::link_model} coupled to an Emax model \eqref{eq::emax}
\begin{eqnarray}
C_\textrm{e}=k_\textrm{e}(C_\textrm{p}-C_\textrm{e}),\label{eq::link_model}\\
E=1-\frac{C_\textrm{e}}{{IC}_{50}-C_\textrm{e}}\label{eq::emax}
\end{eqnarray}
where $C_\textrm{e}$ denotes the effective concentration, $k_\textrm{e}$ denotes the equilibrium constant, $C_\textrm{p}$ denotes the plasma concentration, $E$ denotes the biomarker concentration and ${IC}_{50}$ signifies the concentration required to reach 50\% inhibition  (Emax was assumed to be one). Because of the time delay between concentration and response, it is not intuitively easy to propose sampling time-points that result in informative data. To increase the accuracy of the parameter estimates a preliminary version of PopED lite was used.

The sampling times of the first experiment was optimized based on the model parameters from in vitro data and previously run compounds in the chemical series.  The predicted CV for ${IC}_{50}$ for the optimized design was 25\%.  With the preliminary version of PopED lite, we were only able to conduct Ds optimal design; uncertainty of the initial guess of the parameters was not considered in the design.
The experiment was run with the optimized sampling time and fixed dosage of 0.44 $\mu$mol/kg; however, most of the PD measurements were below the limit of quantification %(see Figure~\ref{fig::CS2_exp1}) 
and ${IC}_{50}$ could therefore not be reliably estimated (CV of ${IC}_{50}$ was over 400\%). The imperfect experimental design was due to a poor estimate of ${IC}_{50}$ based on in vitro data; about one order of magnitude different from that estimated from in vivo data.
It was clear that the dosage should be reduced in the next experiment in order to avoid PD measurements below the limit of quantification; however, it was not clear by how much. 

\textcolor{black}{
In the next experiment, we used a preliminary version of PopED lite to optimize the dose based on the estimated parameters from the first experiment.  The preliminary version of PopED lite had a functionality to generate feasible sets of model parameters (data import followed by bootstrapping) and then run ED(s) optimal design based on these sets.  The preliminary version of PopED lite suggested the new dose to be 0.04 $\mu$mol/kg, and predicted the CV of the ${IC}_{50}$ to be 10\%. The experiment was run with the design suggested by PopED lite and the resulting ${IC}_{50}$ estimate had a CV of 7\%.}% (see Figure~\ref{fig::CS2_exp2}).}

\begin{figure}
\centering
\includegraphics[scale=0.45]{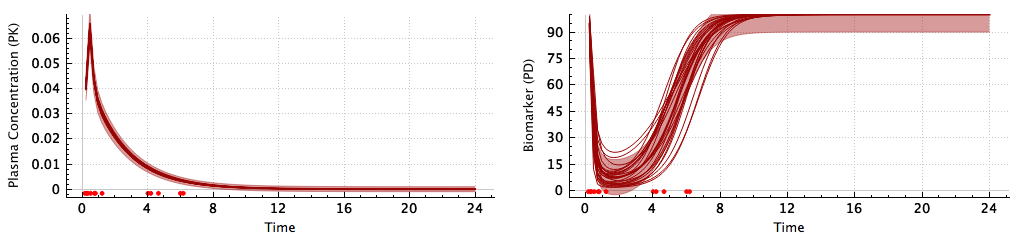}
\caption{Case Study2: Optimized experimental design using the estimated parameters from the first experiment.}
\label{fig::CS2_retro}
\end{figure}

\textcolor{black}{
Similar analysis can be done with the current version of PopED lite, i.e., with 5\% uncertainties in PK parameters and 30\% uncertainties in PD parameters  (Figure~\ref{fig::CS2_retro}).}
%\begin{figure}
%\centering
%\subfigure[Result of  Experiment 1.]{\includegraphics[scale=0.7]{fig_caseStudy2_simulation.pdf}\label{fig::CS2_exp1}}\subfigure[Result of  Experiment 2.]{\includegraphics[scale=0.7]{fig_caseStudy2_simulation2.pdf}\label{fig::CS2_exp2}}\\
%\subfigure[Optimized experimental design using the estimated parameters from the first experiment.]{\includegraphics[scale=0.45]{image8.png}\label{fig::CS2_retro}}
%\caption{Case study 2}\label{fig::study2}
%\end{figure}

\begin{figure}
\includegraphics[scale=0.38]{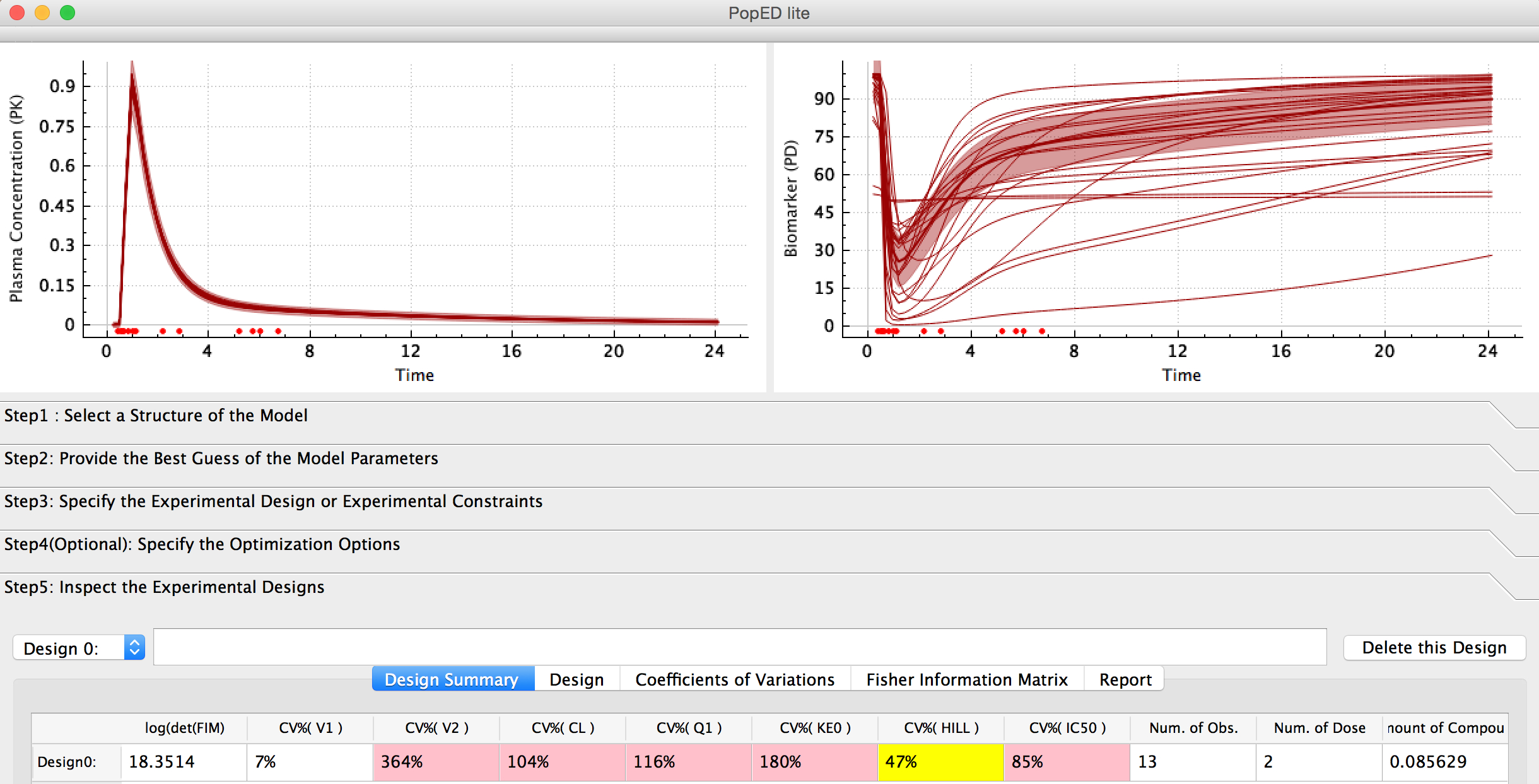}
\caption{Retrospective study of Case study 2}\label{fig::study2_retro}
\end{figure}

As a retrospective study we included 100\% uncertainty to the prior guess of the PD parameters, based on the in vitro data.   As can be seen in Figure~\ref{fig::study2_retro}, using the EDs optimal criterion PopED lite suggested a dose of 0.08 $\mu$mol/kg.  Hence, we could have avoided the uninformative first experiment if the uncertainty of the initial guess of the parameters had been incorporated appropriately in the optimization of the experimental design.  From this case study, we see that a poor initial guess of the parameter values leads to a poor experimental design. To avoid this problem, uncertainty in the prior guess of the parameters should be included in the experimental design calculation.  Based on this case study, we have made ED/EDs optimal design to be the default setting for PopED lite requiring the user to specify the confidence level of the initial guess of the parameters.

\subsection{Case study 3: Designing a mouse study to investigate renal function}
The last case study describes how the deployed version of PopED lite was proven to be a useful tool for motivating an experimental design that was different from the standard design of a preclinical study. The purpose of the study was to investigate renal function in an in vivo mouse model by measuring the glomerular filtration rate (GFR). The standard protocol for GFR measurement is to study clearance of a metabolically inert compound (inulin or sinestrin) that is freely filtered, not bound to plasma proteins and not subject to reabsorption~\cite{Sturgeon1998}. The compound is administered intravenously and the plasma concentration is followed over time. The clearance estimate is subsequently obtained by fitting a standard two compartment PK model to the concentration-curve.

In an initial study, a bolus intravenous injection of sinestrin, and standard sampling-points (3, 7, 10, 15, 35, 55, 75 min) found in literature were used \cite{Qi2004}. However, the compound had an extremely rapid distribution phase in plasma and only approximately 40\% of the mice showed plasma curves that were possible to fit to a two-compartment model.

\begin{figure}
\includegraphics[scale=0.45]{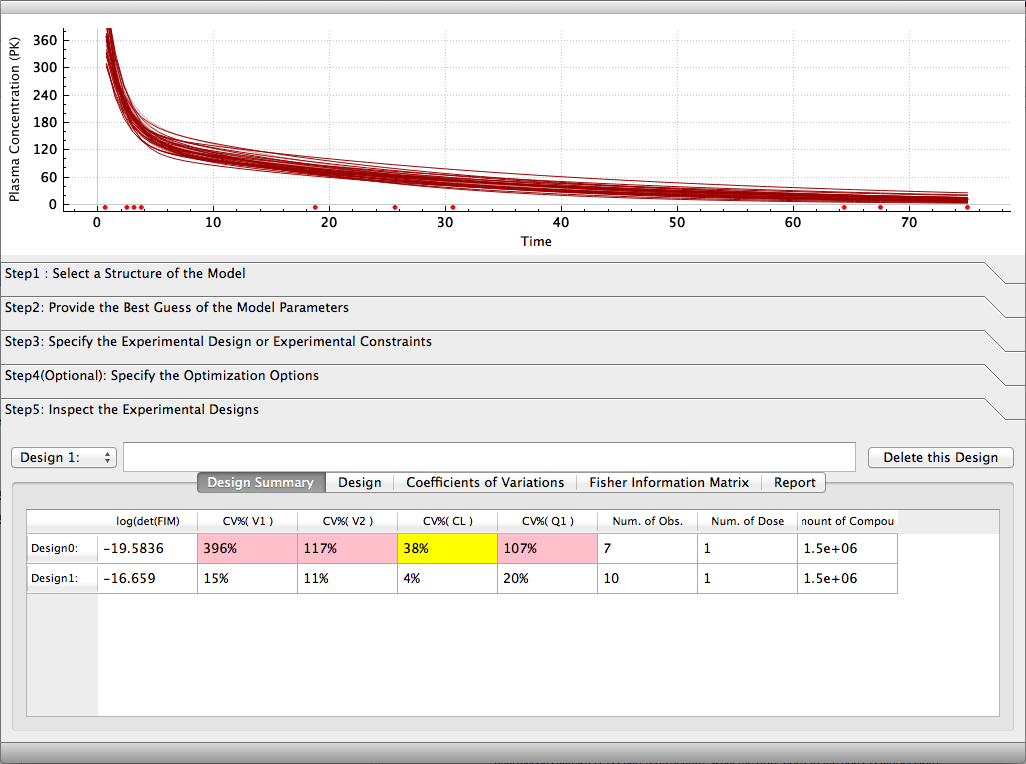}
\caption{Case study 3. PopED lite screenshot of the summary of the experimental designs.  Design0 is the standard design and Design1 is the design optimized by PopED lite.  All CVs are predicted to decrease significantly in the optimized design.
}\label{fig::study4}
\end{figure}

Using PopED lite, the impact of changing the sampling time on the certainty of the estimates could easily be demonstrated (Figure~\ref{fig::study4}). If sampling was restrained to later than 3 minutes the uncertainty in the parameters was much larger than if a sample could be taken at less than one minute. Specifically, we first obtained rough estimates of the PK parameters from previous ran studies and literature values ($V_1$=3000 $\mu l$, $V_2$=5000 $\mu l$, $CL$=300 $\mu l$/min, and $Q_1$=1000 $\mu l$/min), and estimated the uncertainty of these rough estimates to be plus or minus 30\%. Then, we evaluated the standard design by calculating the predicted CVs. The CV of $CL$ was predicted to be approximately 38\% for this study design.  We then calculated the optimal experimental design under the design constraint that seven samples could be taken any minute between 0 min and 75 min. % with dosage of 1 500 000 ng. 
For the new design (1, 2, 5, 10, 35, 75 min), the CV of CL was predicted to reduce to 4\%. Although the proposed design deviated from the standard protocol, the PopED lite calculations convinced the drug discovery team to run the new design.
The resulting measurements from three mice were then modeled with a two-compartment model and the estimated $CL$ had a CV of only 2\%.
Through this successful integration of PopED lite to a preclinical experimental design workflow we showed that even for a very rough initial estimate of the parameters, the experimental design can be improved significantly. %In addition, we have demonstrated that a tool with advanced visualization capability like PopED lite can be a useful tool to convince untraditional designs to experimentalist.

\section{Discussion and Concluding Remarks}
In this paper we have demonstrated how the optimal design software PopED lite was designed collaboratively by academic researchers and preclinical experimental teams in the pharmaceutical industry.  It is designed to fit into a practical workflow, so that the use of an optimal design software can be the standard procedure in preclinical experiment designing and thus help to increase the efficiency of drug discovery in vivo studies.  

Our initial attempt was to introduce the already existing software PopED and create a simplified GUI for preclinical experiments.  However, as we investigated the preclinical experimental design workflow in more detail, we realised that a significant simplification to the user-software interaction, a thorough optimization of the computational speed, and a dramatic reduction of the level of technicality of the software were required.  To address these issues we decided to reconstruct the software from scratch.  This has allowed us to design the software-user interaction first and then implement the numerical computation engines to realise the design.  To optimize the user experience, we have implemented the computation engine with careful treatment of the numerical computational errors, new discrete optimization strategies, and optimized compiled code.

As the result of these optimizations, the optimal design calculation can be done in around ten seconds for models that do not have an ODE and around one minute for  models with stiff ODEs (EDs optimal criteria used and time measured on MacBook Pro Mid 2014 with 2.5GHz Intel Core i7 processr with 16GB of memory).  Future work may include improvement of the computational speed by parallelisation, further  optimisation of the ODE solvers, and refinement of the discrete optimisation algorithm.

%Despite the wide applicability of PopED lite in the preclinical studies, the software is designed specifically for the PK/PD studies with fixed effect models.  We believe a similar software development approach can be beneficial for the studies in the pharmaceutical development where nonlinear-mixed effect models are used.  As the population approach using nonlinear mixed effect models requires more sophisticated FIM calculation and wider flexibility to the possible designs, it may leads us to the further development of the numerical computational methods for the optimal design software.

Deployment of research results through software is not uncommon in the biomedical field and those software have the potential to bridge the gap between theoretically oriented researchers in academia and more practically oriented researchers in the industry.  
%However, theory focused software can often become a high functional ``can do all" software suites that may need specially trained users and not necessary a suitable tool for introducing a new idea.
 The development of PopED lite shows that a cross-disciplinary effort to design a purpose-specific ''app" can be a powerful way of introducing a new research idea to industry.

\section{Acknowledgements}
Authors would like to thank Dr. Sebastian Ueckert and Dr. Joakim Nyberg for valuable and stimulating discussions, Dr. Ron Keizer for sharing his experiences in software development in pharmacometrics and suggesting the Qt framework, and the Uppsala Pharmacometrics Research Group and the AstraZeneca CVMD DMPK modeling and simulation section for useful input and software testing.

\appendix

\section{Fisher Information Matrix} \label{se:appendixFIM}
Let $\boldsymbol{f}(\boldsymbol{\theta};\boldsymbol{\xi})$ be a vector function that maps from the model parameters $\boldsymbol{\theta}$ to the PK and PD prediction of a given experimental design $\boldsymbol{\xi}$. For example, for a standard one-compartment PK model with an $E_{max}$ PD model and an intra-venous bolus administration, the parameter vector is defined as
\begin{eqnarray}
\boldsymbol{\theta}&=&[{V},{CL},{E_{max}},{ED_{50}}]^\textrm{T}.
\end{eqnarray}
The design vector $\boldsymbol{\xi}$ for a design with a single dose and $N_{obs}$ measurement points at times $t_1, t_2,...,t_{N_{\textrm{obs}}}$ is defined as
\begin{eqnarray}
\boldsymbol{\xi}&=&[t_1, t_2, ... , t_{N_{obs}}, \textrm{dose amount}]^\textrm{T}.
\end{eqnarray}
Then, we let the $i$th element of the vector $\boldsymbol{f}(\boldsymbol{\theta};\boldsymbol{\xi})$ be the PK model prediction at time $t_i$ and $(i+{N_{obs}})$th element represent the PD model prediction at time $t_i$ (we assume both PK and PD measurements are made at the same time points).

We define the FIM as
\begin{eqnarray}
FIM(\boldsymbol{\theta};\boldsymbol{\xi})&:=&J^\textrm{T} (\Sigma^{-1}) J\,
\end{eqnarray}
where $J$ is a Jacobian matrix of $\boldsymbol{f}$ with respect to the model parameters $\boldsymbol{\theta}$, i.e.
\begin{eqnarray}
J&=&\left[\begin{array}{cccc}
\die{f_1}{\theta_1}&\die{f_1}{\theta_2}&\cdots &\die{f_1}{\theta_{N_{para}}}\\
\die{f_2}{\theta_1}&\die{f_2}{\theta_2}&\cdots &\die{f_2}{\theta_{N_{para}}}\\
\vdots &\vdots &\ddots &\vdots \\
\die{f_{2N_{obs}}}{\theta_1}&\die{f_{2N_{obs}}}{\theta_2}&\cdots &\die{f_{2N_{obs}}}{\theta_{N_{para}}}\\
\end{array} \label{eq:jacobian}
\right],
\end{eqnarray}
and $\Sigma$ is a diagonal matrix representing the prior knowledge of the measurement error and the limit of quantification.  We interpret the limit of quantification as the minimum value of the residual error, for example, assuming $10 \, \%$ measurement error on the PK observations, $E_{max}/10$ measurement error on the PD observations and 0.01 and 1 to be the limit of quantification of PK and PD measurements (these quantities are specified by the user in Step~2 in the GUI), $\Sigma$ becomes
\begin{eqnarray}
\Sigma&=&\textrm{diag}\left(\left[\max\left(\left|\frac{f_1}{10}\right|,0.01\right)^2,\max\left(\left|\frac{f_2}{10}\right|,0.01 \right)^2,\cdots, \max\left(\left|\frac{f_{N_{obs}}}{10}\right|,0.01\right)^2,\right.\right.\nonumber\\
&& \qquad\qquad\qquad \qquad\left.\left.\max\left( \left|\frac{E_{max}}{10}\right|,1\right)^2,\max\left( \left|\frac{E_{max}}{10}\right|,1\right)^2,\cdots, \max\left(\left|\frac{E_{max}}{10}\right|,1\right)^2 \right]\right).
\end{eqnarray}
Other advanced technique to handle the limit of quantification can be found in \cite{Vong2015} and will be implemented in a future version of the software. 

If $J$ is a full-rank matrix and all the diagonal elements of $\Sigma$ are nonzero, the FIM is a non-singular matrix. In such cases, we define the predicted standard deviation of the parameter estimation uncertainty of the $i$th parameter ($PSD_i$) to be the square root of the $i$th diagonal element of the inverse of the FIM, i.e.
\begin{eqnarray}
PSD_i(\boldsymbol{\theta};\boldsymbol{\xi})&:=& \sqrt{((FIM(\boldsymbol{\theta};\boldsymbol{\xi}))^{-1})_{i,i}}.
\end{eqnarray}
Finally, the predicted CV of the $i$th parameter ($PCV_i$) is defined as
\begin{eqnarray}
PCV_i(\boldsymbol{\theta};\boldsymbol{\xi})&:=&\frac{PSD_i(\boldsymbol{\theta};\boldsymbol{\xi})}{\theta_i} . \label{PCV}
\end{eqnarray}
The predicted CV can be thought of as the predicted accuracy of the estimated parameter from the experiment conducted with the design $\boldsymbol{\xi}$ assuming $\boldsymbol{\theta}$ is the true parameter.

Note that if the Jacobian is a full-rank matrix and the limit of quantification is set to be a positive number, then FIM is a non-singular matrix.  In order to reduce the chance of FIM being numerically singular, we set the lower bound of the limit of quantification to be the square root of the machine accuracy.

%Furthermore, the CV is indicated as inf if the Jacobian matrix $J$ is not a full-rank matrix, e.g. if the observation is not sensitive to some of the parameters. In this way, the user is warned that the model parameters are not identifiable.

Once the FIM is created, we compute the Cholesky decomposition of the FIM.  If the decomposition algorithm breaks down, PopED lite annotates the FIM as numerically singular and outputs a warning that the parameter is not identifiable.  If the decomposition can be computed, PopED lite rebuilds the FIM from the decomposition and compares each element to the corresponding element of the original FIM.  If the maximum absolute value of the difference in an element is more than the square root of the machine accuracy, there is a significant influence of the rounding error in the computation and PopED lite outputs a warning. The decomposition is stored for future usage.

\section{Optimization criterion}
There are several ways to quantify the goodness of the design using the FIM, see e.g. \cite{AtkinsonAC1992, Fedorov1972}. PopED lite implements ED and EDs optimal designs to be the default.
%
%\subsubsection{Design optimization for one parameter}
%One of the main goals in drug discovery is to identify the potency (effect) of drug candidates. Therefore, one important optimal design case is to focus on estimating a specific parameter, typically \(IC_{50}\) or a dissociation constant, with high accuracy. As the predicted coefficient of variation quantifies the predicted accuracy of the parameter estimation, one would like to find a design $\boldsymbol{\xi}^*$ such that
%\begin{eqnarray}
%\boldsymbol{\xi}^*&=&\textrm{argmin}_{\boldsymbol{\xi}}\left( PCV_i(\boldsymbol{\theta};\boldsymbol{\xi}) \right),
%\end{eqnarray}
%where $i$ is the index of the parameter of main interest.  A fundamental assumption of this approach is that the parameter $\boldsymbol{\theta}$ is known. As our goal is to design an experiment in order to increase the accuracy of the parameter(s), needing to have the true parameter to design the experiment is circular logic. On the other hand, it is often the case that we have some rough estimate of the parameter values, i.e., from in vitro experiments or from experiments of a similar compounds. In order to increase the robustness of the experimental design against uncertainty in the prior parameter information, we consider a set of parameters $\{\boldsymbol{\theta}_j \}_{j=1}^{N_{para}}$ that reflects this uncertainty. Then, the optimal design is calculated as
%\begin{eqnarray}
%\boldsymbol{\xi}^*&=&\textrm{argmin}_{\boldsymbol{\xi}}\left( \max_j \left( PCV_i(\boldsymbol{\theta}_j;\boldsymbol{\xi})\right) \right),
%\end{eqnarray}.
%

\subsection{Design optimization for all parameters (ED optimal design)}
Consider the case when the goal of the experiment is to accurately estimate all model parameters. One alternative is to optimize for the sum (or squares of the sum) of the predicted CVs. However, for this approach the optimality of the design can be influenced by the linear reparameterization of the model (see \cite{AtkinsonAC1992} for a detailed discussion). Therefore, we choose instead to quantify the goodness of the experimental design by the determinant of the FIM, which is independent of the linear reparameterization of the model. Taking uncertainty in the prior parameter information into account, the robust optimal design is calculated as
\begin{eqnarray}
\boldsymbol{\xi}^*&=&\textrm{argmin}_{\boldsymbol{\xi}}\left( \sum_{j=1}^{N} \left(\log( \textrm{det}(\textrm{FIM}(\boldsymbol{\theta}_j;\boldsymbol{\xi})))/N\right) \right), \label{eq::ED_criterion}
\end{eqnarray}
where $\boldsymbol{\theta}_j$ are the parameter vectors created uniformly randomly around the ''Guesstimated Parameter Value" with the bound specified as ''Confidence of Guesstimate" (specified in Step~2 of the GUI). Furthermore, $N$ is the number of randomly generated parameter vectors $\boldsymbol{\theta}_j$ (default is set to 25 for practical reasons; however, this setting can be changed in Step~4 of the GUI). 

Note that a naive computation of the determinant of the FIM would often result in accumulation of rounding errors. In PopED lite the log of the determinant of the FIM is calculated by summing the log of the diagonal elements of its Cholesky decomposition and multiplying the sum by two.
If Cholesky decomposition cannot be computed due to a singularity of the FIM, the log of the determinant of such matrices will be set to $-10^{10}$ and not to negative infinity. In this way, PopED lite selects the experimental design with the least possibility of an unidentifiable system.

As alternatives to \eqref{eq::ED_criterion}, the following robust criteria are implemented in PopED lite (can be set in Step~4 in the GUI):

\begin{eqnarray}
\boldsymbol{\xi}^*&=&\textrm{argmin}_{\boldsymbol{\xi}}\left( \min_{j} \left(\log( \textrm{det}(\textrm{FIM}(\boldsymbol{\theta}_j;\boldsymbol{\xi})))\right) \right),\\
\boldsymbol{\xi}^*&=&\textrm{argmin}_{\boldsymbol{\xi}}\left( \textrm{median}_{j} \left(\log( \textrm{det}(\textrm{FIM}(\boldsymbol{\theta}_j;\boldsymbol{\xi})))\right) \right).
\end{eqnarray}

\subsection{Design optimization for a subset of parameters (EDs optimal design)}
As a second option, consider the case when a given subset of the parameters is of particular interest. The robust optimal design is then calculated as
\begin{eqnarray}
\boldsymbol{\xi}^*&=&\textrm{argmin}_{\boldsymbol{\xi}}\left(  \sum_{j=1}^{N}\log \left( \frac{\textrm{det}(\textrm{FIM}(\boldsymbol{\theta}_j;\boldsymbol{\xi}))}{\textrm{det}(\textrm{FIM}_{\backslash subset}(\boldsymbol{\theta}_j;\boldsymbol{\xi}))}\right)/N \right), \label{eq::EDs_critarion}
\end{eqnarray}
\textcolor{black}{where $\backslash subset$ denotes the complement of the subset of the parameters of particular interest (i.e., subset of parameters of no-interest).}
If only one parameter is chosen to be of interest, the above optimization criterion reduces to optimizing for the CV for that particular parameter. EDs optimal design is used when the options "Optimize the Design for the parameter estimation accuracy of" is set to be either ''PK parameters only" or ''PD parameters only" in Step~2 in the GUI. As an alternative to Eq.~\eqref{eq::EDs_critarion}, the following criterion can be optionally used in PopED lite (Step~4 in the GUI):
\begin{eqnarray}
\boldsymbol{\xi}^*&=&\textrm{argmin}_{\boldsymbol{\xi}}\left(\sum_{j=1}^{N}\log \left( \textrm{det}(\textrm{FIM}_{subset}(\boldsymbol{\theta}_j;\boldsymbol{\xi}))\right)/N \right).
\end{eqnarray}

\section{Weighted Random Search Algorithms} \label{se:optAlgorithm}
In this appendix, we describe the discrete optimization algorithms we have implemented in PopED lite.  To avoid local convergence and to take advantage of the discrete nature of the optimization problem, we have constructed stochastic algorithms to find an approximately optimal design in a reasonable computational time.
These algorithms are made specifically to address the discrete constrained optimisation problem that appeared as a result of the design of PopED lite.

\subsection{Sampling time optimization}
Consider the problem of choosing $N_\textrm{obs}$ observation points from $N_\textrm{disc}$ discrete possible observation points. For example, if $N_\textrm{obs}=10$ and $N_\textrm{disc}=120$ there are over $10^{14}$ possible combinations and an exhaustive search is not computationally feasible. Roughly speaking, our algorithm aims to randomly search for the best design; however, for each iteration the probability of the randomly chosen observation points are adjusted based on the previous evaluation of the experimental designs.  The key feature of this weighted random search algorithm is that each individual weight is defined by the combination of a pair of possible observation points, and that all weights then can be stored in a matrix (i.e., the $i$th row $j$th column of the weight matrix represents the probability of choosing $i$th possible observation point given $j$th possible observation point is already chosen to be an observation point).

 Let $\phi(D)$ be a scalar function that takes a subset of natural numbers and returns a scalar value that we aim to maximise (e.g., the subset of the natural numbers corresponds to the discrete time points in the possible observation time window and the scalar value corresponds to the determinant of the Fisher Information Matrix), i.e.
\begin{eqnarray}
\phi : \mathbb{D} \rightarrow \mathbb{R},
\end{eqnarray}
where $\mathbb{D}$ is the space of all the subsets of the natural numbers between 1 and $N_\textrm{disc}$ whose size is $N_\textrm{obs}$. Note that size of $\mathbb{D}$ is $N_\textrm{disc}!/((N_\textrm{disc}-N_\textrm{obs})!N_\textrm{obs}!)$ which grows rapidly as $N_\textrm{disc}$ increases when $N_\textrm{obs}$ is fixed.

All the vectors that can potentially appear as a row of the Jacobian \eqref{eq:jacobian}, for any design, that is used to construct the FIM can be precomputed before the start of the sampling time optimization.  Hence the evaluation of each experimental design does not require any PK/PD model simulation during the optimization and is hence not very computationally intensive.  As a result, the computational cost of storing and checking already evaluated designs exceeds the computational cost of the redundant evaluation of the same design. Thus the weighted random search algorithm for the sampling time optmization allows the redundant evaluation of the same design.

The algorithm is defined as follows.
%(an element of this matrix $a_{ji}$ represents the un-normalized probability of the $j$th index chosen given the $i$th index has been chosen previously).\\
\\
\textbf{Pseudocode}\\
 Construct an $N_\textrm{disc} \times N_\textrm{disc}$ matrix $A$ with all elements assigned to 1. (This matrix $A$ stores the pairwise weights for the random search and updated after each iteration.)\\
For $m=1,2,...,N_\textrm{iter}$\\
\indent For $l=1,2,...,N_\textrm{test}$\\
\indent \indent  Reset the temporary weight matrix $B$ to be the weight matrix $A$ and empty the set $D_l$, i.e.,
\begin{eqnarray}
B&=&A\\
D_l&=&\emptyset.
\end{eqnarray}
\indent \indent  Uniformly randomly choose a natural number between $1$ and $N_\textrm{disc}$ and let it be $i$ and add it to the set $D_l$.\\
\indent \indent  Set the $i$th row of the temporary weight matrix $B$ to be 0, i.e.,
\begin{eqnarray}
 b_{ij}=0\qquad \textrm{for all }j=1,2,...,N_\textrm{disc}. \label{eq:no_re_samp1}
\end{eqnarray}
\indent \indent  For $n=2,3,...,N_\textrm{obs}$\\
\indent \indent\indent  Randomly choose a natural number between $1$ and $N_\textrm{disc}$ with the probability
\begin{eqnarray}
\frac{b_{ki}}{\sum_{k=1}^{N_\textrm{disc}}b_{ki}}\qquad \textrm{for } k=1,2,...,N_\textrm{disc}, \label{eq::wightedProbComp}
\end{eqnarray}
\indent \indent\indent  and let it be $i$ and add it to the set $D_l$.\\
\indent \indent\indent  Set the $i$th row of the temporary weight matrix $B$ to be 0 so that the $i$th sampling time will not be\\
\indent \indent\indent  chosen again, i.e.,
\begin{eqnarray}
 b_{ij}=0\qquad \textrm{for all }j=1,2,...,N_\textrm{disc}. \label{eq:no_re_samp2}
\end{eqnarray}
\indent \indent  End for\\
\indent End for\\
 \indent  Find $D^*_m$ such that
\begin{eqnarray}
\phi(D^*_m)=\max_{l=1,2,...,N_\textrm{test}}\phi(D_l)\,.
\end{eqnarray}
 \indent  Update the weight matrix $A$ as follows
\begin{eqnarray}
a_{ij}=a_{ij}+1 \qquad \textrm{for all } (i, j)\in D^*_m \times D^*_m. \label{eq:updateWeight}
\end{eqnarray}
End for\\
Return the best subset $D^*$ such that
\begin{eqnarray}
\phi(D^*)=\max_{m=1,2,...,N_\textrm{iter}}\phi(D_m^*)\,.
\end{eqnarray}

The search algorithm assumes that multiple observations at the same point are not allowed. Consequently, the probability of selecting an already chosen point is assigned zero at \eqref{eq:no_re_samp1} and \eqref{eq:no_re_samp2}. If multiple sampling for the same time point is allowed then simply omit \eqref{eq:no_re_samp1} and \eqref{eq:no_re_samp2} from the algorithm. If sampling from two consecutive discrete possible observation points is not allowed (e.g. since the animal needs to rest between sampling points) the rows adjacent to the $i$th row can be set to zero as well.

Note that the algorithm reduces to a pure random search if the weight matrix is not updated at \eqref{eq:updateWeight}. As a default, PopED lite uses the weighted random search algorithm for sampling time optimization without possibility of multiple sampling from the same time point.

Further extension of the algorithm such as defining the weights as a tensor so that the weight given the combination of more than two observation points can be stored is possible.  %Also Equation~\eqref{eq::wightedProbComp} can be modified as 
%\begin{eqnarray}
%\sum_{j\in D_l}\frac{b_{kj}}{\sum_{k=1}^{N_\textrm{disc}}b_{ki}}\qquad \textrm{for } k=1,2,...,N_\textrm{disc}.
%\end{eqnarray}

\subsection{Dosage optimization}
Consider the problem where $N_\textrm{doses}$ doses are given and each dose needs to be chosen between $d_\textrm{min}$ and $d_\textrm{max}$.  We first discretize the possible dose range into  $N_\textrm{pd}$ possible doses $\{d_1, d_2, d_3, ... d_{N_\textrm{pd}} \}$, where $d_i\neq d_j$ if $i\neq j$ and $d_\textrm{min}\leq d_i\leq d_\textrm{max}$.  We now wish to obtain a vector of $N_\textrm{doses}$ indices $[i,j,k,l,...]$  to maximise the optimization criterion when dose amount $d_i$ is given at the first dose event, $d_j$ is given at the second dose event, etc.  Note that there are $(N_\textrm{pd})^{N_\textrm{doses}}$ possible dose combinations and that the number grows rapidly as $N_\textrm{doses}$ increases.  It is therefore often not feasible to conduct an exhaustive search for a multiple dose experiment.  Instead, we randomly search through the possible dose combinations and select the best dose combination.  To save computation time, this weighted random search algorithm is designed to avoid evaluation of previously evaluated designs.

Note that this optimization problem is fundamentally different from the sampling-time optimization problem, as for the sampling time optimization, we aim to find the optimal subset of indices (i.e., the order of the indices does not matter) while for the dosage optimization we aim to find a vector of indices (i.e., the order of the indices matters).  

\begin{figure}
\centering
\includegraphics[bb=0 0 778 291, scale=0.62]{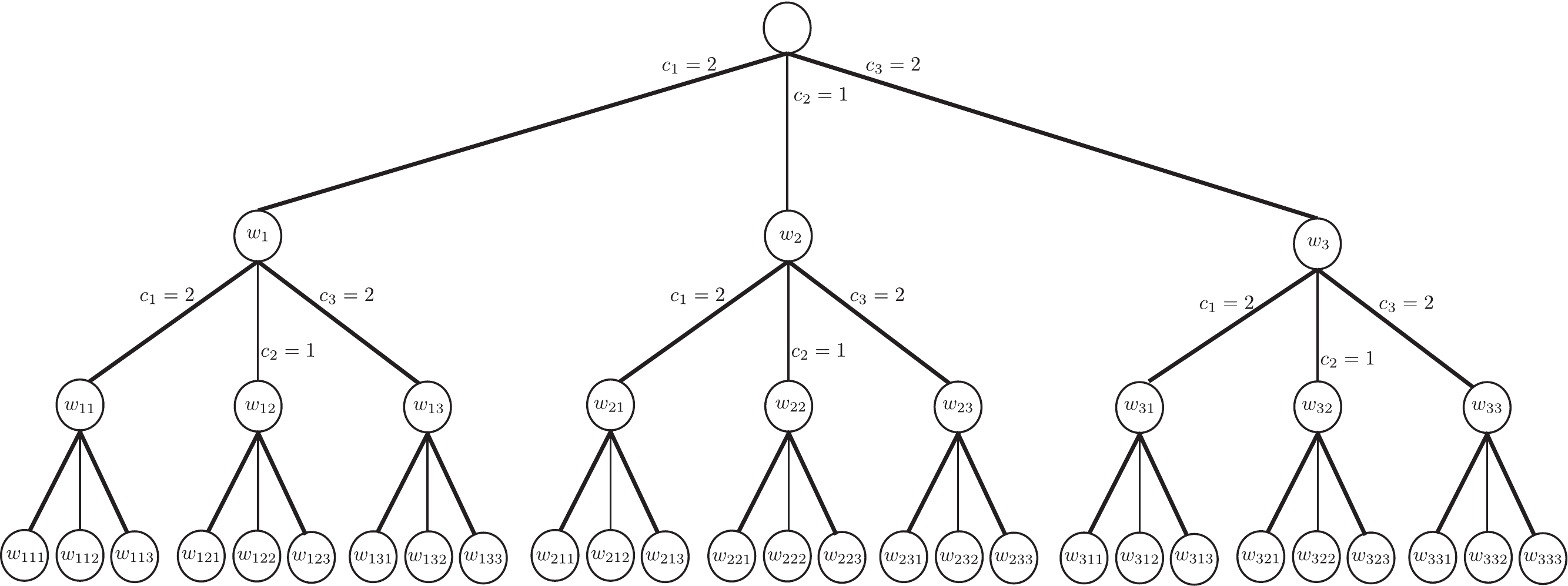}
\caption{An example of a tree for storing the weight for the weighted random search for dosage optimisation ($N_\textrm{pd}=2$, $N_\textrm{doses}=3$).  In this example, the highest and lowest possible dosages have priorities on the search and the design vectors $[111],[113],[131],[133],[311],[313],[331],[333]$ are more likely to be evaluated before other design vectors.}\label{fig::tree}
\end{figure}

For this dosage optimization algorithm, we construct an ordered tree to store the weights.  This tree is constructed by nodes storing the weights $w$ and edges storing weight biases $c$.  Each path of the tree represents the design of an experiment, for example the path reaches from the root to a leaf following $i$, $j$,$k$th edges represents the experimental design vector $[i,j,k]$ (i.e., $d_i$ as the first dose, $d_j$ as the second dose and $d_k$ as the last dose).  An example of the tree is depicted in Figure~\ref{fig::tree}.

Weight bias $c$ define the fundamental importance of the possible dosage that is represented as edges in the tree.  For example, if there are 10 possible dose levels for each dosing time, it is wise to search for a reasonably good combination of high, medium, or low doses in a first step, and then fine tune the dose levels in a second step.  Hence, in this case we put more weight bias to these three doses, and lower weight bias to the rest.  By choosing the weight bias $c$ this way, the algorithm will most probably first search through the optimal combination between high, medium and low doses and then conduct the fine tuning during the remaining search. Weight bias $c$ remain the same throughout the search.

The variable $w$ is a weight of the probability of choosing the dosage that is represented by the edge connected to the parent node.  For example, $w_{ij}$ (the weight stored in the node that can be reached by path $i,j$ from the root node) is the weight for the random choice where $j$th possible dosage should be chosen as a second dosage given $i$th possible dosage is chosen for the first dosage.  The weight $w$ depends on the weights of all the descendent nodes.  
For the nodes that are leaves the weights $w$ are initially defined as weight bias $c$ of the connecting edge and after each evaluation of the design, the weight of the corresponding leaf is reduced to zero to avoid the redundant evaluation of the same design.  Hence the weights $w$ are recalculated after each design evaluation.

We describe the algorithm more in detail using an example with $N_\textrm{doses}=3$.  The extension to other cases is trivial.

%The tree for the case of $N_\textrm{pd}=2$ can be depicted as follows:\\
%\Tree[. [.$(w_1,c_1)$ [.$(w_{11},c_{11})$ [.$(w_{111},c_{111})$ ] [.$(w_{112},c_{112})$ ]]
%               [.$(w_{12},c_{12})$ [.$(w_{121},c_{121})$ ] [.$(w_{122},c_{122})$ ]]]
%          [.$(w_2,c_2)$ [.$(w_{21},c_{21})$ [.$(w_{211},c_{211})$ ][.$(w_{212},c_{212})$ ]]
%                [.$(w_{22},c_{22})$ [.$(w_{221},c_{221})$ ][.$(w_{222},c_{222})$ ]]]]\\
%                \\

The weights of the nodes are defined as follows:
\begin{eqnarray}
w_i&=&c_i\sum_{j=1}^{N_\textrm{pd}}w_{ij} \label{eq:tree1}\\
w_{ij}&=&c_{j}\sum_{k=1}^{N_\textrm{pd}}w_{ijk} \label{eq:tree2}\\
w_{ijk}&=&c_{k} \qquad \textrm{ for all }i,j,k.
\end{eqnarray}
We consider the case where we randomly choose $N_{\textrm{search}}$ dose combinations and pick the best dose combination among them.\\
\\
\textbf{Pseudocode}\\
For $m=1,2,...,N_\textrm{search}$
\begin{itemize}
\item Randomly choose an integer $l$ between $1$ and $N_\textrm{pd}$ with the probability $ w_l/\sum_{i=1}^{N_\textrm{pd}}w_i$, and let $i=l$.
\item   Randomly choose an integer $l$ between $1$ and $N_\textrm{pd}$ with the probability $ w_{il}/\sum_{j=1}^{N_\textrm{pd}}w_{ij}$, and let $j=l$.
\item   Randomly choose an integer $l$ between $1$ and $N_\textrm{pd}$ with the probability $ w_{ijl}/\sum_{k=1}^{N_\textrm{pd}}w_{ijk}$, and let $k=l$.
\item  Evaluate the design with dose amount $d_i$ at the first dose time, dose amount $d_j$ at the second dose time, and  dose amount $d_k$ at the third dose time.  If this design is better than the one found previously keep this design, else discard.
\item  Let $w_{ijk}=0$ so that the dose combination of $ijk$ will not be redundantly evaluated during the following search. 
\item  Recalculate the weights following \eqref{eq:tree1}-\eqref{eq:tree2}.
\end{itemize}
End for\\
Return the best design.\\

Note that, the weight calculation needs to be implemented recursively so that the weight can be calculated for any $N_\textrm{doses}$.
%In order to keep the flexibility for $N_\textrm{doses}$, the weights need to be calculated in a recursive manner.  
In addition, to reduce the computational cost of constructing the tree, we construct only the necessary part of the tree to calculate the new weights after each search iteration.  Note that this algorithm avoids the redundant evaluation of the design; hence if $N_\textrm{search}$ is sufficiently large for given $N_\textrm{doses}$ and $N_\textrm{pd}$, the search is exhaustive.  In addition, avoiding redundant evaluation of the same design is crucial for the dosage optimisation as each evaluation of the design requires PK/PD model evaluation and can be computationally intensive.

The weight bias $c$ of this algorithm is set to one in the current version of PopED lite.  Future work may include changing $c$ dynamically throughout the search.

\subsection{Simultaneous optimization of dose and sampling time}
To optimize both dose and sampling time, conduct sampling time optimization at each search iteration in the dose optimization.

%When the optimization of both dose and sampling time are the interest of the user, at each search iteration in the dose optimization, conduct sampling time optimization to evaluate the best sampling time for the chosen dose combination. 

%The algorithm above assumes that multiple observations at the same point is not allowed hence set the probability of choosing the already chosen point to be zero at \eqref{eq:no_re_samp1} and \eqref{eq:no_re_samp2}.  However, if multiple sampling of the same point is allowed then omit \eqref{eq:no_re_samp1} and \eqref{eq:no_re_samp2} from the algorithm.  Also, if for example, sampling from two consecutive discrete possible observation points is not allowed (e.g., need to have some time for the animal to rest between the sampling of the blood) then we can set the rows adjacent to $i$th row to be also zero, i.e., replace \eqref{eq:no_re_samp1} and \eqref{eq:no_re_samp2} with the following expression:
%\begin{eqnarray}
% b_{kj}=0\qquad \textrm{for all }\left\{\begin{array}{l}k=i-1,i,i+1, \\ j=1,2,...,N_\textrm{disc}.\end{array}\right. \label{eq:no_re_samp3}
%\end{eqnarray}
%In addition, one can note that if we do not update the weight matrix at \eqref{eq:updateWeight} then it will be the pure random search.  As a default, PopED lite uses the weighted random search algorithm for sampling time optimization without 

\section{Matrix/Vector Notations}
In this paper we denote a scalar quantity with a lower case letter e.g., $a$ or $c$, a matrix with a capital letter, for example $A$, or $M$, and a vector quantity by a bold symbol of a lower case letter, e.g., $\boldsymbol{v}$, $\boldsymbol{a}$, unless otherwise specifically stated.
As an exception to this matrix notation, Fisher Information Matrix is denoted as FIM to follow the convention of the notation in the field.
In addition the $i$th row of the matrix $M$ can be denoted by a vector $\boldsymbol{m}_{i \cdot}$, the $j$th row of the matrix $M$ can be denoted by a vector $\boldsymbol{m}_{\cdot j}$ and the $i$th row $j$th column of the matrix $M$ can be written by a scalar $m_{ij}$.

\end{doublespace}
\bibliography{popedlite}{}
\bibliographystyle{plain}
\end{document}